\documentclass[preprint,showpacs,preprintnumbers,amsmath,amssymb]{revtex4-1}
\usepackage{graphicx}  
\usepackage{epsfig}
\usepackage{subfigure}
\usepackage{bm}        
\usepackage{amssymb}   
\usepackage{color}
\usepackage{amsmath}
\def\e{\begin{equation}}
\def\f{\end{equation}}
\def\-#1{{\bf #1}}

\def\va{\varepsilon}

\def\.{\cdot}

\def\##1{{\bf#1\mit}}

\def\r#1{(\ref{eq:#1})}

\begin{document}

\title{Enhanced Efficiency of Light-Trapping Nanoantenna Arrays for Thin Film Solar Cells}
\author{Constantin R. Simovski$^1$, Dmitry K. Morits$^1$, \\
Pavel M. Voroshilov$^2$, Michael E. Guzhva$^2$, \\
Pavel A. Belov$^2$, Yuri S. Kivshar$^{2,3}$}
\affiliation{$^1$Department of Radio Science and Engineering,
Aalto
University, Aalto FI-00076, Finland\\
$^2$National Research University of Information Technologies,
Mechanics and Optics (ITMO), St. Petersburg 197101, Russia\\
$^3$Nonlinear Physics Center, Research School of Physics and
Engineering, Australian National University, Canberra ACT 0200,
Australia}


\begin{abstract}

We suggest a novel concept of efficient light-trapping structures
for thin-film solar cells based on arrays of planar nanoantennas
operating far from plasmonic resonances. The operation
principle of our structures relies on the excitation of chessboard-like collective
modes of the nanoantenna arrays with the field localized between
the neighboring metal elements. We demonstrated theoretically
substantial enhancement of solar-cell short-circuit current by the
designed light-trapping structure in the whole spectrum range of
the solar-cell operation compared to conventional structures employing
anti-reflecting coating. Our approach provides a general
background for a design of different types of efficient broadband
light-trapping structures for thin-film solar-cell technologically
compatible with large-area thin-film fabrication techniques.

\medskip


\end{abstract}

\maketitle

\section{Introduction}

The extensive study of thin-film solar cells (TFSC) aims for the
production of large-area panels harvesting the solar energy.
Prospective TFSC imply their implementation on flexible substrates
compatible with the so-called roll-to-roll
processing~\cite{nanosolar}. Such solar cells are characterized by
a very small amount of purified semiconductor per unit
area~\cite{SC1} which reduces the cost being useful for
ecology~\cite{SC1,SC2}.

To make efficient solar cells of a very small thickness,
an anti-reflecting coating should be replaced by
{\em a light-trapping structure}, since none of the conventional
anti-reflecting coatings can prevent the transmission of light
through a photovoltaic layer. This transmission results in energy
loss and substrate heating, which lead to an additional reduction
 of the solar cell efficiency~\cite{SC1,SC2}. A light-trapping structure
(LTS) is a structure capable to reduce both the reflection from
a solar cell and transmission through its photo-absorbing layer.

Many suggestions for a LTS design are based on photonic
crystals (e.g., Refs.~\cite{SC3,SC4,SC5}), nano-textured
semiconductor coatings (e.g., Refs.~\cite{SC6,SC7,SC8}), and plasmonic
nanostructures (e.g., Refs.~\cite{SC9,SC10,SC11}). Some LTSs performed
as regular plasmonic gratings~\cite{SC11,SC12,Spinelli} operate
similarly to the semiconductor textured coatings, i.e. they
convert incident plane waves into waveguide modes propagating inside the
photovoltaic layer. The operation of textured coatings obeys
certain basic restrictions~\cite{SC12,SC13,SC14} known as the
Yablonovich limit. It relates the field confinement inside a
semiconductor slab with the maximal operation band and minimal
optical thickness of the slab. This limit, however, refers to the
ray optics, and in the case of plasmonic gratings it can be
overcame by involving surface plasmon polaritons~\cite{SC12,Spinelli,SC20}.

Many ideas of the efficient light trapping for TFSC are based on
the use of plasmonic absorbers~\cite{SC13,SC14,SC9,SC10,SC11}
which are planar arrays of silver or gold nanoparticles with
plasmon resonances within the operation band of solar cells.
Plasmonic absorbers can be divided into two classes. The first
class is random arrays of nanoparticles where only
averaged parameters (averaged size of the particles and their surface
density) are optimized. Such structures are fabricated by
self-assembly, and they have the geometry close to that of a
pillar or a tablet.  These plasmonic absorbers can be engineered
so that at certain frequencies both reflection from the solar cell
and transmission through the photovoltaic layer are strongly suppressed
\cite{SC14,SC9,SC10,SC11}. From a macroscopic point of view, such an
absorber can be treated as an effective lossy layer with a
broadband resonance of the complex permittivity. The general
drawback of these absorbers are high losses in metal
nanoparticles and noticeable scattering.

The second class of plasmonic absorbers is regular grids of
silver or gold nanoparticles located on a top of traditional
vertical p-n or p-i-n junctions (e.g., Refs.~\cite{SC15,SC16,SC17,SC18}).
We call such structures arrays of nanoantennas. Such nanoantenna arrays
can demonstrate several plasmon resonances~\cite{SC15,SC18}, and they
can operate even as dielectric optical cavities \cite{SC19}. Nanoantennas
not only offer a significant enhancement of the photovoltaic absorption,
but they can concentrate the field outside the metal elements avoiding
 useless dissipation of the solar energy.

In this paper, we suggest and study theoretically a novel type of
light-trapping structures based on arrays of nanoantennas excited
far from plasmonic resonances.  We demonstrate that such
non-resonant absorbers can enhance substantially the
photoabsorption in very thin (100-150 nm) semiconductor layers and
increase significantly the overall spectral efficiency of solar
cells with a very small thickness of an active layer. Our LTS are
in-plane optically isotropic (being also independent on the light
polarization) coatings located on the top of a solar cell. We
discuss possible implementations of our concept in the solar cell
technology and analyze in detail two specific examples of TFSC:
inter-band TFSC based on copper-indium-gallium-selenide (CIGS) and
silicon TFSC.

\section{Physics of enhanced light trapping}

\begin{figure}[h]
\epsfig{file=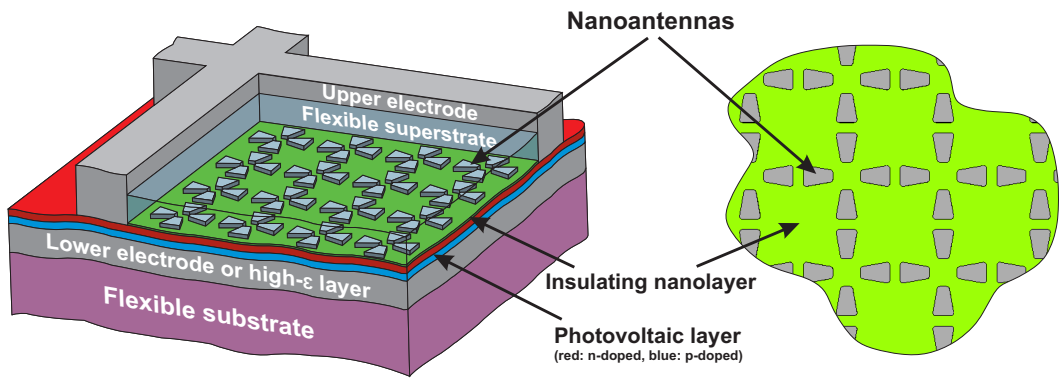, width=14cm} \caption{(Color online)
A schematic of thin-film solar cell with a light-trapping structure
(left) and a top view of the nanoantenna arrays (right).}
 \label{Pavel1}
 \end{figure}

A schematic view of our LTS is shown in Fig.~\ref{Pavel1}. This design
employs the advantages of collective oscillations excited in the visible
or infrared spectral range by an incident plane wave in a lattice of
Ag nanobar nanoantennas.  The chessboard-like modes in our structure
are analogous to the collective oscillations discussed in
Ref.~\cite{domino} for arrays of parallel metal bar operating in the far-infrared
range and termed~ \emph{domino modes}. Such modes exist are excited at several
 wavelengths  by the external electric field whose vector is orthogonal to the
 bar axis, and they are characterized by the domination of higher multipole
 moments induced in every nanobar over its dipole moment. The main feature
 of such modes is the advantageous distribution of the local electric
 field which is concentrated in the gaps between the nanobars. The internal field
in metal is small, and consequently these modes demonstrate low losses.
The chessboard-like modes can be excited also in the visible range~\cite{Olli}
if the thickness of the bar is within 50-100 nm and its length is larger
than 200-250 nm.

In several earlier studies~\cite{domino,Olli}, these modes were excited
by an incoming beam with the grazing incidence on the array plane.
However, our simulations reveal that these modes are also excited
for any angle of incidence (e.g. by a normally incident wave) as
well as in the presence of a substrate or without any substrate.

When these modes are excited in arrays of nanobar nanoantennas
located in free space~\cite{Olli}, the absorption maxima at the
wavelength of these modes are weak. Their excitation implies the
electromagnetic energy is stored between the nanobars without an
impact on reflectance and transmittance. This situation changes
dramatically if the nanobars are located in the absorbing
host or on an absorbing substrate. Then the electromagnetic energy
of the near-field concentrated in between the nanobars is absorbed
by a host medium. In this way both reflection and transmission of
the incident wave can be suppressed at several wavelengths.

The presence of a substrate leads to a change of the bar
cross section compared to that discussed in Ref.~\cite{Olli}.
For arrays of nanobars located on a semiconductor substrate,
the domino modes correspond to a rectangular cross section,
whereas the bar thickness should be 2-3 times as smaller
as the width. In our simulations the parallel nanobars
located on a-Si substrate have the thickness 40-70 nm, width
100-150 nm, and length 200-500 nm. These domino modes are excited at
several wavelengths in the visible range, and their total band has
a relative width close to 20\%. The further enlargement of
the band became possible by employing tapering of nanobars. This
design allows overlapping of separate wavelength ranges of
the domino modes which form a solid band with a relative width 40-50\%.
This band is still narrower than the operational band of a
practical solar cell, and nanobars of LTS should not worsen
the operation of the solar cell beyond the band of the light
trapping. This means that the arrays of tapered nanobars should
be sparse enough to prevent strong reflection from metal
elements.

\begin{figure}[!h]
\epsfig{file=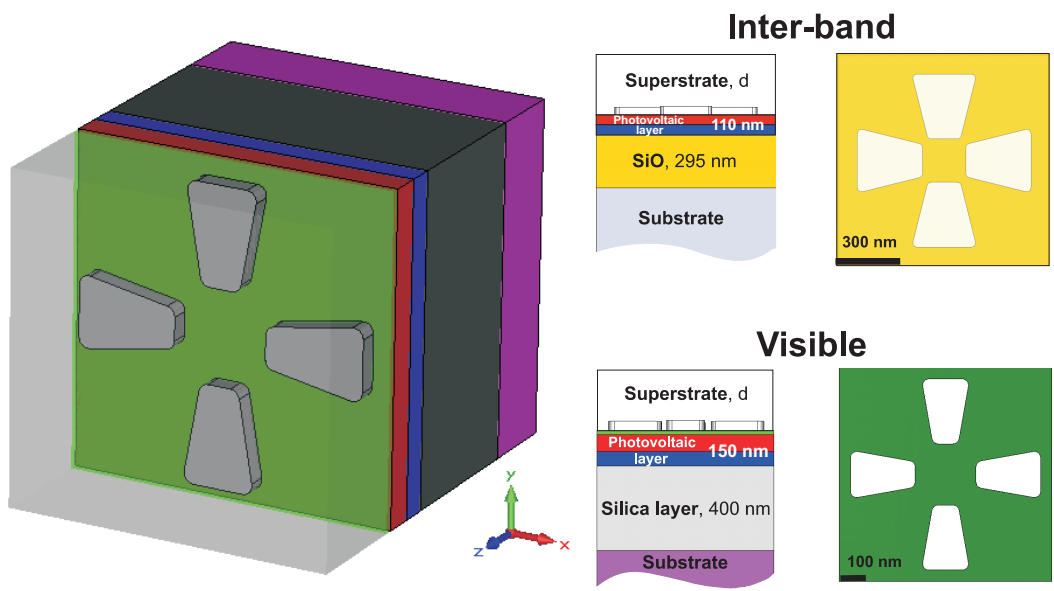, width=14cm} \caption{(Color
online) Left: A unit cell of the structure under consideration.
Left: details of the explicit design. Top: CIGS-based inter-band
solar cell, bottom: Si-based solar cell for the visible range.
Side view and top view of the unit cell are given in scale with
the reference length unit. Different colors of p-doped  and
n-doped parts of the Si layer reflect different levels of optical
losses due to different concentration of carriers. In the
case of CIGS, both p- and n-doping levels are equivalent.}
 \label{Morits1}
 \end{figure}

Next, in order to achieve the polarization-independent
response, our LTS should comprise of at least two orthogonal arrays
of light-trapping nanobars. Two latter requirements result in a
unit cell performed as a cross of tapered nanobars. The unit
cell shown in Fig.~\ref{Morits1} operates as a broadband
polarization-independent nanoantenna. However, light trapping
in the arrays is a collective effect which disappears when the gap
between the adjacent nanoantennas becomes too large.

The underlying physical mechanism of light trapping due to the excitation of low-loss
chessboard-like domino modes can be complemented by the effect of the Fabry-Perot cavity
 which can be engineered in the adjacent frequency range. We managed to achieve
this regime in our numerical simulations for the case when an additional dielectric layer
of high permittivity (of the order of 3-4) is introduced between the substrate
and  photovoltaic layer. For the substrate of amorphous silicon, it can be a layer
of silicon monoxide. Then, in a desired frequency range, the whole structure
creates a Fabry-Perot cavity with both small transmission and small reflection
of incident waves.

\section{Suggested design and numerical results}

We calculate numerically the power absorbed by a unit cell of the
TFSC structure with LTS within the operation band $\Delta
\omega=[\omega_1,\omega_2]$ of the photovoltaic layer. The power $A(\omega)$
absorbed in the photovoltaic layer is defined as an integral of the local
power density over the volume $V$ of the layer per unit cell, and
the power $A_m(\omega)$ absorbed in the nanoantenna is defined as the integral
over the volume of metal per unit cell. In both the cases, the
local electric field is calculated for the normally incident plane
wave with the unit amplitude of the electric field. Our calculations are
performed for two cases: The first case is LTS and a blooming layer,
the latter is a simplest type of ARC, and it is used
for comparison. Though anti-reflecting properties of multilayer
anti-reflection coating are slightly better, they are achieved by
 a nanometer precision being more sensitive to nanometer
 variations of the thickness than the reflectance of a single blooming layer
\cite{Shalin1,Shalin2}. As a result, the optical properties of a
multilayer ARC suffer of abrasion more than those of a simple ARC.
Therefore, we compare the short-circuit current of a TFSC
calculated in the presence of our LTS and that of a single
dielectric layer whose permittivity and thickness are optimized
for minimizing the reflection.

Our numerical simulations are carried out using both
Ansoft HFSS and CST Studio software packages. The meshing is
refined so that the result would not depend on the computational
grid, and the convergence is carefully controlled. The agreement
between the results obtained with the HFSS and CST Studio packages
is excellent: the plots presented below visually coincide.
Moreover, we employ two options of the Ansoft HFSS package -- an
incident plane wave and a wave port -- and obtained the same
coincidence. The unit cell of the LTS shown in Fig.~\ref{Morits1}
refers to the TFSC region located in the gaps between the
wires of the electrodes collecting the photo current. These mesh
electrodes should be located on both top and bottom of the
photovoltaic layer. We do not consider the regions shaded by the
wire mesh assuming that they are much smaller than the regions
open to the sunlight.

First, we show that the gain in the short-circuit current due to the
presence of the LTS equals to the gain in the absorption of the
incident wave with uniform spectrum in the photovoltaic layer.
This absorption should be averaged over the operation band
$[\omega_1,\omega_2]$ of the solar cell with a certain weight function
$f(\omega)$ as determined below. The short-circuit current per unit
area of the solar cell $J_{sc}$ is equal to (see e.g.
\cite{SC10,SC15}): \e
J_{sc}=\int\limits_{\omega_1}^{\omega_2}A_p(\omega)R_s(\omega)d\omega,\quad
A_p(\omega)={\omega\va_0\va''(\omega)\over 2}\int_{V}|E(\omega,\-r)|^2dV. \label{f1}\f
Here $R_s(\omega)$ is the spectral response of the photovoltaic
material whose complex permittivity has the imaginary part $\va''$,
and $\-E(\omega,\-r)$ is the electric field produced by the solar
radiation inside the photovoltaic layer. Since the problem is
linear, the field $\-E(\omega,\-r)$ can be expressed through the solar irradiance
$I_s(\omega)$ as $|\-E(\omega,\-r)|^2={I_s(\omega)}|\-E_n(\omega,\-r)|^2$, where
$\-E_n(\omega,\-r)$ is the field produced by the incident wave of unit
amplitude at every frequency. Then we can write
$A_p(\omega)=A(\omega)I_s(\omega)$, where \e A(\omega)={\omega\va_0\va''\over
2}\int_{V}|E_n(\omega,\-r)|^2dV, \label{aaa}\f is the spectral
photo-absorption of the wave with unit amplitude. We can further
express the value $J_{sc}$ in the form: \e J_{sc}=\int\limits_{\omega_1}^{\omega_2}
I_s(\omega)R_s(\omega)A(\omega)d\omega\equiv \Delta \omega <A>. \label{f2}\f Here $<A>$ is
the spectral absorption averaged over the band $\Delta
\omega=\omega_2-\omega_1$ with the weight function $f(\omega)\equiv
 I_s(\omega)R_s(\omega)$ which is a product of two table values (photovoltaic spectral
response of the photovoltaic material layer and the spectrum of
solar irradiance at the earth surface). To calculate the
short-circuit current, we simulate the wave absorption
with an unit amplitude over the band $[\omega_1,\omega_2]$, and then
average it with the weight function $f(\omega)$. After calculating
$<A>$ for the case when the TFSC is covered by LTS and the
case when it is covered by a dielectric ARC,  we obtain
the gain due to the replacement of the ARC by LTS:  \e G\equiv
{J_{sc}^{LTS}\over J_{sc}^{ARC}}={<A>^{LTS}\over <A>^{ARC}}.
\label{gain}\f

The first photovoltaic structure for which we perform the
optimization of the nanoantenna arrays is a 110-nm thick n-p layer
of CuInSe$_2$ located between a polyethylene superstrate and a
layer of silicon monoxide. This kind of CIGS (when Ga is fully
substituted by In) corresponds to a solar cell with photovoltaic
spectral response and optical losses essentially overlapping in
the inter-band region 250-450 THz~\cite{CIGS0,CIGS1}. The
thickness of the CuInSe$_2$ layer is chosen as small as 110 nm in
order to stress the light-trapping capacity of our LTS. From other
point, this thickness is larger than the depletion region around
the p-n junction and smaller than the minimum carrier diffusion
length~\cite{CIGS_L,CIGS_L1}. Therefore, the selected thickness of
the photovoltaic layer corresponds to a ratio between
the photo-current and photo-absorption~\cite{SC2}. In this
regime, the increase of the photovoltaic absorption has the maximal
possible impact on the overall efficiency of the solar cell.

The optical constants of doped CuInSe$_2$ are taken from
Ref.~\cite{CIGS}, and the optical constants of polyethylene are
 taken from Ref.~\cite{Polyethilene}, for the region 400-450 THz,
and Ref.~\cite{Polyethilene_IR}, for the region 250-400 THz. Those
of the silicon monoxide are taken from Ref.~\cite{silica}.
The optimal thickness of the polyethylene layer is $d=$270 nm
whereas it is 295 nm for SiO. Other parameters are shown in
Fig.~\ref{Morits1}.

\begin{figure}[!h]
\subfigure[]{\includegraphics[width=0.57\linewidth]{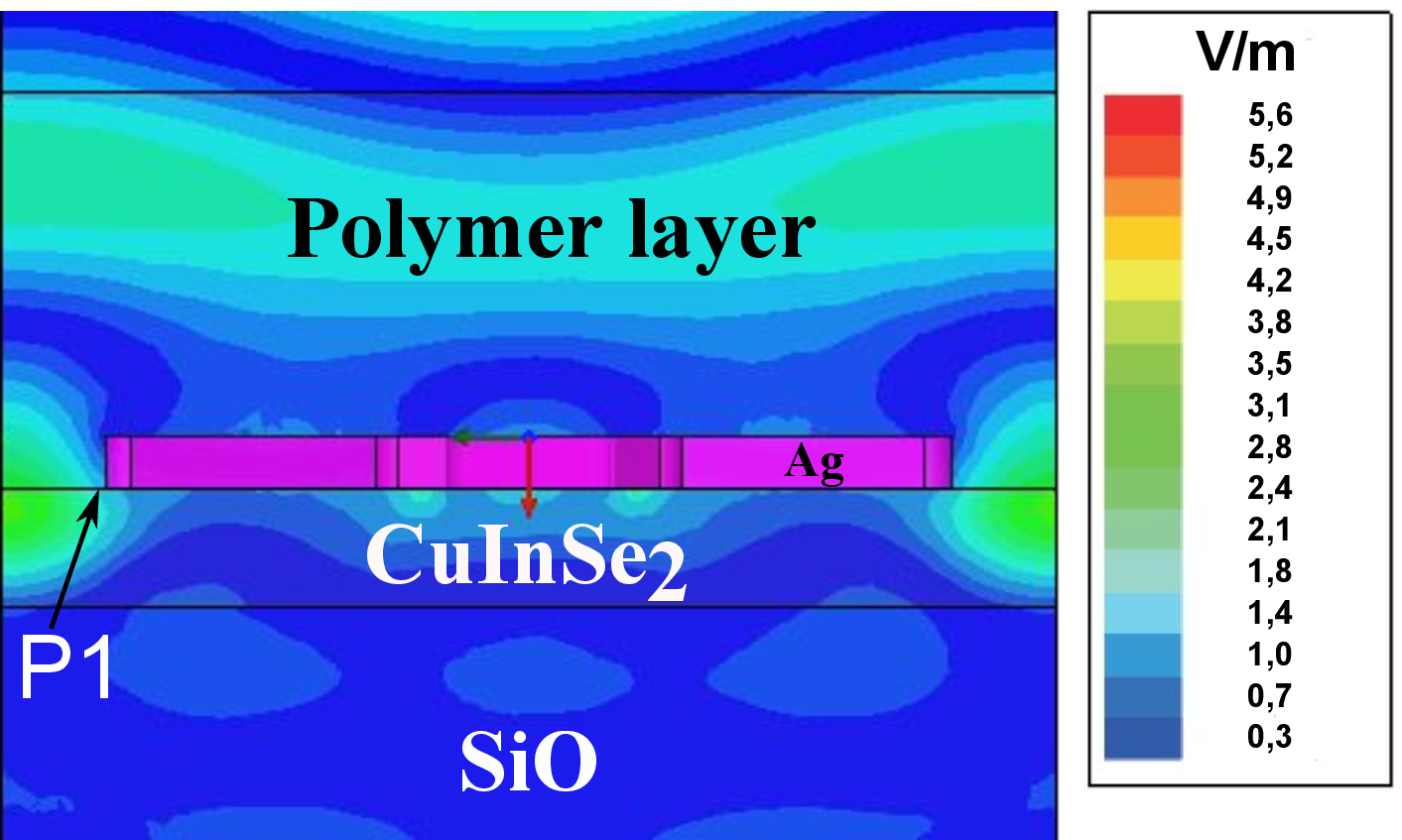}}%
\subfigure[]{\includegraphics[width=0.43\linewidth]{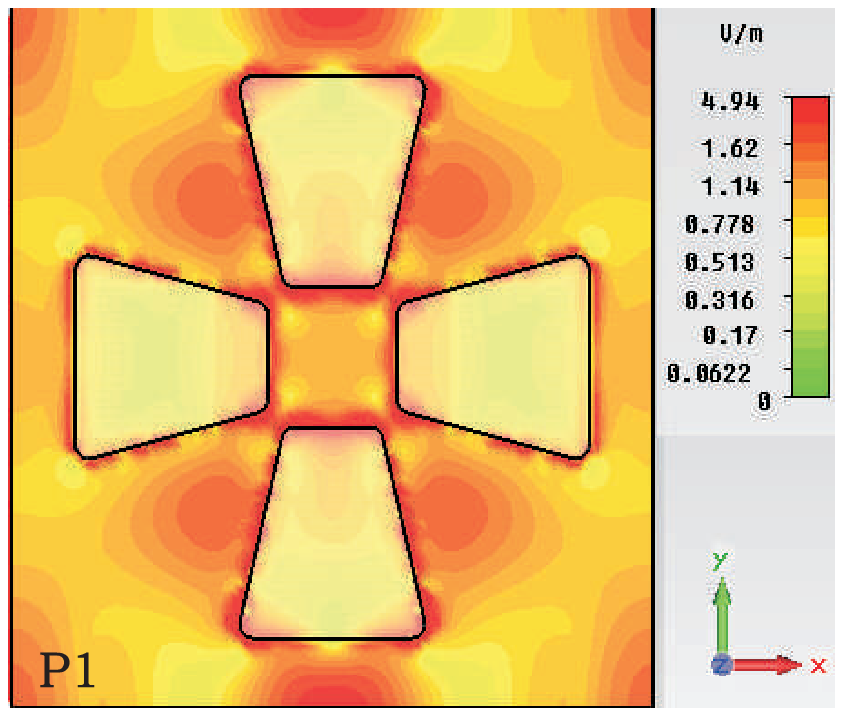}}
\caption{(Color online)  Electric field amplitude for 370 THz
illustrating the concept of the LTS: (a) central vertical cross section;
(b) horizontal plane P1 (upper interface of CuInSe$_2$).
The incident wave has the amplitude of 1 V/m.}
 \label{Morits2}
 \end{figure}

Within the frequency range 350-450 THz, the light is trapped by nanoantenna
arrays where the chessboard-like domino modes are excited. This is seen from the local
field distributions over the unit cell at these frequencies. In
Figs.~\ref{Morits2}(a,b), we show the local field distribution at
the frequency 370 THz in the central vertical cross section of the
unit cell and in the horizontal plane P1 -- upper surface of the
insulating layer of amorphous silica separating nanoantennas from
the semiconductor. We vary its thickness within the range 2-20
nm and find that its presence affects none of the main
characteristics, such as absorption in the semiconductor and
metal elements, and the coefficients $R$ and $T$. In
Fig.~\ref{Morits2}(a) we observe a rather strong reflected field.
Due to the presence of nanoantennas, the reflectance increases
nearly by 30\%. However, the decrease of the transmittance is more
important, and the light-trapping effect is quite significant.
Inspecting the field distribution in Fig.~\ref{Morits1}(a), we observe
a practical absence of the transmitted field. In both the cases shown in
Figs.~\ref{Morits1}(a,b) we find that the field is concentrated beyond the
metal elements, and the harmful dissipation of the solar power
inside the LTS is thus prevented.

Location of the local field maxima in Fig.~\ref{Morits1} outside
the photovoltaic layer (mainly in the dielectric superstrate) is
not a drawback of our LTS design. To maximize the photo-absorption, it is
not mandatory to concentrate the field inside the photovoltaic
layer.  Evanescent field concentrated in the dielectric
superstrate is neither absorbed or nor radiated. The absorption
occurs only in those parts of hot spots which are located inside
the metal and semiconductor. It is important that the useful
absorption $A$ in the semiconductor exceeds strongly the harmful
absorption $A_m$ in the metal.

\begin{figure}[!h]
\includegraphics[width=0.6\linewidth]{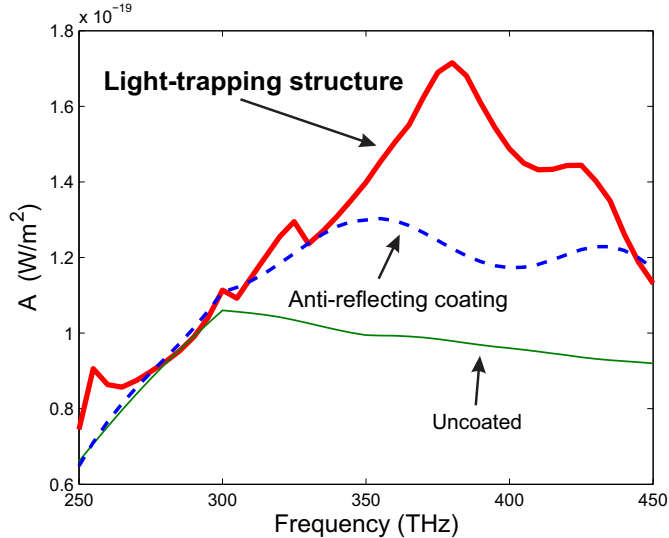}
\caption{(Color online) Absorption in the inter-band
photovoltaic layer for three cases: (i) solar cell enhanced by
LTS, (ii) solar cell enhanced by ARC, and (iii) uncoated solar cell.}
 \label{Morits3}
 \end{figure}

In the frequency range 250-350 THz, the nanoantennas are weakly excited and
the light-trapping occurs due to the regime of the Fabry-Perot
cavity. In Fig.~\ref{Morits3}, we show the spectral photo-absorption
$A(\omega)$ for three cases: with our LTS, with a blooming layer, and
without any structure on the top of the photovoltaic layer. The
optimal blooming is achieved using the same polyethylene film
with the thickness $d=$440 nm. We notice that the refraction index of
the polyethylene $n_{PE}$ corresponds approximately to the
blooming condition for the half-space of CIGS ${\rm
Re}(n_{CIGS})\approx n_{PE}^2$ over a significant part of the
operation band.

The reflection losses averaged over the operation frequency
band in the presence of LTS are found to be close to 20\%,
averaged transmission losses are about 8\%, while only
6\% of the incident power is absorbed in the LTS nanoantenna arrays.
This gives us an useful absorption $<A>$ about 66\% of the incident power.
with the presence of ARC, the reflection losses are 7\% and transmission
losses are as large as 46\%. IN this later case the absorption equals
 49\% (cf. with 38\% in the non-coated case). The gain due to the LTS is estimated
as $G=$1.34 compared to the case of ARC. Using formula
\r{gain}, we obtain $G=$1.29 due to the presence of the weight
function $f(\omega)$ in the averaged absorption $<A>$.  Compared to
the non-coated cell, the LTS gives the gain of $G=$1.7.

We believe that the obtained enhancement is sufficient to justify
the fabrication costs of our LTS nanoantennas. Below, we discuss
the possibility to fabricate our LTS in a way which seems to be
rather inexpensive under the condition of the mass production and
compatibility with the roll-to-roll processing~\cite{nanosolar}.

Our second numerical example of LTS corresponds to the silicon solar
cell whose operation band is practically coincide with the range of
the visible light. The study of the second design aims to demonstrate
 that our design concept allows different types of TFSC operating in
different frequency bands. For this latter case, the dimensions shown
in Fig.~\ref{Morits1} are numerically optimized for the
photo-absorbing layer of the thickness 150 nm. The density of carriers
in the layer is assumed to be $3\cdot 10^{18}$ cm$^{-3}$.
The dielectric superstrate is amorphous silica as well as the ARC
in the reference solar cell. The simplistic blooming condition
${\rm Re}(n_{Si})\approx n_{\rm silica}^2$ does not provide a good
estimate for  a multi-layer structure with a high optical contrast
of layers. A silica film with the thickness $d=$175 nm reduces
the reflection averaged over the visible range almost 5 times
(from 45\% to 10\%) enhancing the averaged absorption by 40\%.

\begin{figure}[!h]
\includegraphics[width=0.6\linewidth]{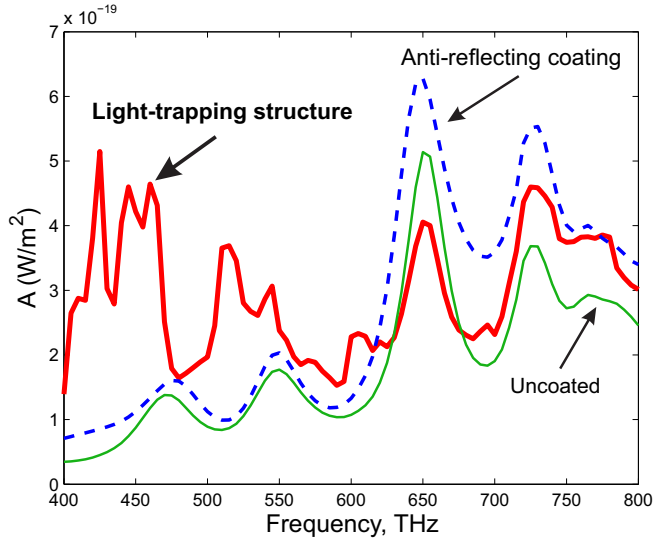}
\caption{(Color online) Absorption in the silicon photovoltaic
layer for three cases: (i) solar cell enhanced by LTS, (ii) solar cell
enhanced by ARC, and (iii) uncoated solar cell.}
 \label{Voroshilov2}
 \end{figure}

In Fig.~\ref{Voroshilov2}, we show the frequency dependence of the
photovoltaic absorption $A(\omega)$ for three three cases discussed
above: (i) optimal LTS, (ii) optimal blooming layer, and (iii) non-coated
solar cell. The best superstrate thickness for our LTS is equal to
$d=$325 nm, however we find that the absorption  depends weakly on
the thickness $d$ in the interval $d=$325-390 nm, and the nanometer precision is
not required. In the presence of nanoantennas, the light-trapping effect
related to the domino modes is observed in the region 400-600 THz. At
600-800 THz the nanoantennas are weakly excited and only slightly
worsen the blooming operation of the silica superstrate. The gain
in the photovoltaic absorption calculated in accord with Eq.~\r{f2} is
17\% ($G=1.17$). Compared to the non-coated TFSC, our LTS
gives the gain 64\%. These results are close to those obtained for
the CIGS solar cells.

Choosing the best material for nanoantennas, we study
different metals and find that only silver can properly
satisfy the light-trapping functionality. With other metals
 the domino modes are not found. We believe that for other metals either
  absolute value of the complex permittivity is not sufficiently large (Au)
or optical losses are too high (Cr, Al, Cu).

\section{Fabrication issues}

Now we discuss briefly possible technologies for the fabrication of
suggested arrays of nanoantennas with a dielectric superstrate
in a way compatible with the roll-to-roll processing~\cite{nanosolar}
for large-area TFSC panels. First, we suggest to fabricate the arrays
of nanoantennas on a polymer film using the replication technology~\cite{Novac}
the latter implies a quartz template whose surface repeats the profile of
the nanoantenna arrays. Using the method described in Ref.~\cite{Novac}
 with this template, one can obtain a vast amount of plastic replicas
with metal nanoantennas reproduced with practically the same resolution
as that of the template. Besides polyethylene, these plastic substrates
can be made of polyamide, polymethylmetaacrylate, polytetrafluorethylene,
polystyrene, etc. Pieces of the film (of size 0.5-1 mm) with printed
nanoantenna arrays can be prepared separately from  solar cells so that
the price of every piece will be small. Then these pieces can be placed
on the top of the solar cell in the gaps between the wires of the mesh electrode.
If it is done in a vacuum camera and under mechanical pressure, a strong bonding
arises between the polymer and the silica covering the surface of
the solar cell. In accord to our estimates, this bonding would be
sufficient for adhesion, and the nanoantennas will be fixed on the solar cells
as well as the plastic superstrate. Finally, the thickness of the
superstrate can be reduced chemically to a few submicrons. This
procedure is feasible if the gaps between the wires of the contact
mesh are as large as 0.5-1 mm which are typical values for known
mesh electrodes of TFSC structures~\cite{Meier}.

Alternatively, our structures can be fabricated by using conducting polymers
\cite{anodes} or any other flexible transparent electrode. Then
the nanoantennas can be fabricated on the top of a solar cell
protected by a silica insulator being covered by a flexible
conducting coating. This coating will play twofold role: as an
upper electrode and also a superstrate of our light-trapping structure.
To prevent an ohmic contact with the superstrate, the nanoantenna arrays
 can be covered with organic molecules or a nanolayer of deposited silica.
In the gaps between nanoantennas, the silica insulator should be
removed from the surface of the photovoltaic layer. In both the cases
our design is compatible with the concept of the roll-to-roll
processing of thin-film solar cells.

\section{Conclusions}

We have suggested and analyzed theoretically a novel design of
light-trapping solar-cell nanopatterned structures which allows a significant
enhancement of the photovoltaic absorption in the layers as thin
as 110 nm. Our design is material-independent, and it can be
applied to a variety of solar cells operating in the visible, inter-band or
infrared frequency range. More importantly, the light-trapping
functionality of our structures is broadband, and it is not based
on spectrally narrow plasmonic or any other resonances. Light trapping
in the structures originates from the excitation of chessboard-like modes
(also called domino modes) of a lattice of tapered silver nanobar
nanoantennas which can be excited for different geometric
parameters within a vast frequency spectrum ranging from far-IR
to the visible.  We have demonstrated that our nanoantenna
arrays operate significantly better than the structures based on
anti-reflecting coatings. We have discussed the design issues
with respect to possible mass-fabrication of our nanoantennas.

\section{Acknowledgements}

This work has been supported by the Ministry of Education and
Science of Russian Federation, the Dynasty Foundation and the Australian Research
Council. The authors thank K.R. Catchpole and H. Savin for useful
discussions and suggestions.

\end{document}